\documentclass[vecphys]{svmult}

\usepackage{makeidx}         
\usepackage{graphicx}        
\usepackage{multicol}        
\usepackage[bottom]{footmisc}

\makeindex             

\begin{document}
\def\aj{AJ}%
\def\araa{ARA\&A}%
\def\apj{ApJ}%
\def\apjl{ApJ}%
\def\apjs{ApJS}%
\def\ao{Appl.~Opt.}%
\def\apss{Ap\&SS}%
\def\aap{A\&A}%
\def\aapr{A\&A~Rev.}%
\def\aaps{A\&AS}%
\def\azh{AZh}%
\def\baas{BAAS}%
\def\jrasc{JRASC}%
\def\memras{MmRAS}%
\def\mnras{MNRAS}%
\def\pra{Phys.~Rev.~A}%
\def\prb{Phys.~Rev.~B}%
\def\prc{Phys.~Rev.~C}%
\def\prd{Phys.~Rev.~D}%
\def\pre{Phys.~Rev.~E}%
\def\prl{Phys.~Rev.~Lett.}%
\def\pasp{PASP}%
\def\pasj{PASJ}%
\def\qjras{QJRAS}%
\def\skytel{S\&T}%
\def\solphys{Sol.~Phys.}%
\def\sovast{Soviet~Ast.}%
\def\ssr{Space~Sci.~Rev.}%
\def\zap{ZAp}%
\def\nat{Nature}%
\def\iaucirc{IAU~Circ.}%
\def\aplett{Astrophys.~Lett.}%
\def\apspr{Astrophys.~Space~Phys.~Res.}%
\def\bain{Bull.~Astron.~Inst.~Netherlands}%
\def\fcp{Fund.~Cosmic~Phys.}%
\def\gca{Geochim.~Cosmochim.~Acta}%
\def\grl{Geophys.~Res.~Lett.}%
\def\jcp{J.~Chem.~Phys.}%
\def\jgr{J.~Geophys.~Res.}%
\def\jqsrt{J.~Quant.~Spec.~Radiat.~Transf.}%
\def\memsai{Mem.~Soc.~Astron.~Italiana}%
\def\nphysa{Nucl.~Phys.~A}%
\def\physrep{Phys.~Rep.}%
\def\physscr{Phys.~Scr}%
\def\planss{Planet.~Space~Sci.}%
\def\procspie{Proc.~SPIE}%
\let\astap=\aap
\let\apjlett=\apjl
\let\apjsupp=\apjs
\let\applopt=\ao

\title*{(Sub)mm Interferometry Applications in Star Formation Research}
\author{Henrik Beuther\inst{1}}
\institute{Max-Planck-Institute for Astronomy, K\"onigstuhl 17, 69117 Heidelberg, Germany
\texttt{beuther@mpia.de}}
%
%

\maketitle


\section{Introduction}
\label{sec:1}

Interferometry at (sub)mm wavelengths is one of the most important
tools to study the physical and chemical properties of the youngest
and most embedded stages of star formation. High spatial resolution is
needed to resolve small-scale structure, e.g., details of accretion
disks, molecular outflow or the planet formation processes in low-mass
star formation. Furthermore, in high-mass star formation research high
spatial resolution is similarly important because the regions are on
average more distant, and massive star formation proceeds always in a
clustered mode. The (sub)mm wavelengths bands are important because
young and deeply embedded stages of star formation are characterized
by cold gas and dust temperatures of the order 10\,K. At such low
temperatures, the peak of the Planck-blackbody curve is at
$\lambda_{\rm{max}}[\rm{mm}]=2.9/T=290\mu$m (Wien's law) and one can
easily observe the whole Rayleigh-Jeans tail at (sub)mm wavelengths.
Therefore, the combined requirement of high spatial resolution and
(sub)mm wavelengths focus the research interest to (sub)mm
interferometry.

While early exploratory interferometry at even longer wavelengths
started already shortly after World War II, real imaging
interferometry rather began at cm wavelengths in the 1970s with
instruments like the the Very Large Array (VLA) in the USA and the
Westerbork Synthesis Array in the Netherlands (in the 1960s). During
the 1980s, technical progress allowed to go to even shorter
wavelengths, and mm facilities like the Plateau de Bure Interferometer
(PdBI) in Europe, the Berkeley-Illinois-Maryland-Array (BIMA) and the
Owens Valley Radio Observatory (OVRO) in the USA, and the Nobeyama
Millimeter Array (NMA) in Japan were built to explore the mm
wavelength range at high angular resolution. Only a few years ago, in
2003, the Submillimeter Array (SMA) managed to reach the submm window
with interferometric techniques, and the Atacama Large Millimeter
Array (ALMA) will cover all earth-accessible (sub)mm wavelengths
windows from the next decade onwards.

Each (sub)mm and cm wavelengths window has its own advantages and
disadvantages, and one has to select the right regime according to the
scientific questions. For example, at cm wavelengths, the free-free
emission is strong, one finds important maser transitions (e.g.,
H$_2$O, CH$_3$OH, OH, etc.), a few interesting molecules have easy
accessible spectral lines there (e.g., OH, NH$_3$), and one can
observe the 21\,cm line of neutral hydrogen there. Going to the mm
regime, the dust continuum emission rises strongly, and one can
observe the dense dust and gas cores. Furthermore, the mm regime
harbors the low-energy lines of one of the most important
astrophysical molecules Carbon Monoxide (CO $J=1-0,\,2-1$). In addition
to CO, many more complex molecules have important transitions in the
mm regime, and one can trace various physical and chemical properties
with the large range of molecular line transitions. Stepping to the
submm window, the dust and line emission rises even more strongly
($S_{\rm{cont}}\propto \nu^{\alpha}$ with $2<\alpha<4$,
$S_{\rm{line}}\propto \nu^5$).  Regarding the temperature and density
regimes many molecules are sensitive for, the mm regime rather favors
spectral lines emitted from the colder gas ($10<T<50$\,K), whereas one
finds in the submm regime mainly lines from warmer and denser gas
($T>50$\,K). However, this differentiation is only on average valid,
since one finds extremely highly excited lines already at cm
wavelengths (e.g., high $(J,K)$ NH$_3$ inversion lines), and other
low-level lines can be observed in the submm regime (e.g., some
CH$_3$OH lines).

The scientific questions possible to target with (sub)mm
interferometry in star formation research cover nearly all physical
and chemical processes one may think of. Here, I will present some
of the most important applications in the framework of this European
network: obviously molecular outflows, then the closely related
accretion disks, furthermore the question of fragmentation and early
sub-structure formation, as well as astrochemical applications. Last
but not least, I will give an outlook of the future (sub)mm array
ALMA.

\section{Molecular outflow}
\label{outflows}

\paragraph{Low-mass outflows}

Shocks associated with molecular outflows and jets have first been
detected in the 1950s \cite{herbig1951,haro1952}. Later in the 1970s,
first non-Gaussian line-wing emission from molecular CO was detected
\cite{kwan1976}. While molecular outflows were originally not
predicted by theory, after their observational detection it was
obvious that they are a crucial ingredient to remove angular momentum
during star formation. Today, it is well established that star-forming
regions of all masses and luminosities drive molecular outflows,
however, it is still debated whether the outflow driving mechanisms
are always the same (e.g., \cite{arce2006}). In low-mass star
formation the current paradigm includes the formation of a
centrifugally supported accretion disk and associated outflows/jets
driven by magneto-centrifugal acceleration (e.g., \cite{pudritz2006}).

The text-book example of a low-mass outflow is the one observed toward
the class 0 source HH211. The region was observed in CO(1--0) with the
PdBI covering both outflow lobes in a large mosaic \cite{gueth1999}.
They found a collimated jet-like component at velocities $>10$\,km/s
with respect to the $v_{\rm{lsr}}$, and a less collimated, spatially
broader component at lower velocities. This morphology was interpreted
in the framework of jet-entrainment where a collimated high-velocity
jet entrains additional gas from the surrounding envelope.  More
recent follow-up observations in several transitions of SiO and higher
excited CO lines revealed the jet-like nature of this outflow in even
more detail (Fig.~\ref{palau}, \cite{hirano2006,palau2006}).
Corresponding SiO knots are closer to the driving source in more
highly excited lines indicating higher temperatures at the knot shock
fronts. Furthermore, SiO(8--7)/(5--4) line ratios decrease with
distance from the driving source which is indicative of a density
gradient in the surround gas core. Last but not least, it was found
that the SiO emission exhibits always higher velocities than CO at the
same projected position of the molecular outflow (Fig.~\ref{palau},
\cite{palau2006}). This is interpreted again in the jet-entrainment
picture where the SiO emission is closer to the axis of the primary
beam and hence at higher velocities whereas the CO emission traces
entrained gas further outside. However, one has to keep in mind that
even such jet-like molecular components are not the primary jet
ejected by the protostar-disk system, but that even the observed
jet-like molecular gas is already entrained gas. Observed optical jets
show far higher velocities of the order a few 100\,km/s
\cite{mundt1990}. Different jet-entrainment processes have been
proposed, e.g., entrainment via bow-shocks at the head of the jet,
turbulent entrainment via Kelvin-Helmholtz instabilities or wide-angle
winds, for more details see a recent discussion of the various
entrainment possibilities \cite{arce2006}.

\begin{figure}[htb]
\centering
\caption{The {\bf left} panel presents the recent SMA observations of
the molecular jet observed at submm wavelengths in the SiO(8--7) and
CO(3--2) lines \cite{palau2006}. Blue and red contours show the line
emission as labeled in each panel and the grey-scale outlines the
shocked H$_2$ emission. The grey-contours in the middle panel outline
NH$_3$ emission, and the SiO(8--7) spectrum in the bottom panel is
averaged over the central $10''$ of the jet. The {\bf right} panel
shows the corresponding position-velocity diagram with SiO(8--7) in
grey-scale and CO(3--2) in contours.}
\label{palau}       
\end{figure}

While the jet-entrainment scenario is relatively well established
today for low-mass star formation, other interesting topics are under
discussion. For example, optical slit spectroscopy observations toward
DG Tau found rotation of the optical jet around its outflow axis
\cite{bacciotti2002}, and complementary mm interferometer
investigations revealed the corresponding accretion disk around the
protostar with the same rotation orientation as the jet
\cite{testi2002}.  This is strong support for the tight disk-jet
connection and that the jets have to be launched from the
protostar-disk interface. The main two competing theories for the
jet-launching are magneto-centrifugal disk winds emanated from an
extended inner disk region (e.g., \cite{pudritz2006}) and the X-wind
theory predicting that the jet-launching region is close to the inner
disk truncation radius, at the so-called X-point \cite{shu2000}. Ray
et al.~(2006) \cite{ray2006} presented new jet-rotation observations
that did not actually trace the disk-launching region, but that
allowed to extrapolate at what distance from the protostar the disk is
most likely to be launched.  Although these observations are not
finally conclusive, they are consistent with current disk-wind models
favoring a more extended disk-region as jet-launching site.

\paragraph{High-mass outflows}

The picture of molecular outflows from high-mass star-forming regions
has changed considerably over the last few years mainly based on
interferometric observations at mm wavelengths. Since the mid-90s,
increasing evidence arose that molecular outflows are ubiquitous
phenomena in high-mass star formation as well (e.g.,
\cite{shepherd1996a,beuther2002b}). However, early single-dish mapping
studies of high-mass outflows claimed that the collimation degree of
massive outflows is lower than known for low-mass outflows (e.g.,
\cite{shepherd1996b,richer2000}). This was interpreted as support for
alternative formation scenarios going to higher-mass stars (e.g.,
\cite{bonnell1998,stahler2000}). In contrast to this claim,
single-dish observations with better angular resolution revealed that
the observed collimation degrees of massive outflows are consistent
with those from their low-mass counterparts as soon as the larger
distances and lower spatial resolution of such regions are properly
taken into account \cite{beuther2002b}. As the next step,
interferometric observations of some high-mass star-forming regions
clearly resolved the previously chaoticly appearing single-dish
observations into multiple outflow systems from various members of the
evolving high-mass protocluster (e.g.,
\cite{beuther2002d,beuther2003a,gibb2003,su2004,garay2003,davis2004}).
Figure \ref{05358} shows the multiple collimated outflows in the
massive star-forming region IRAS\,05358+3543 as an example. These
observations support a picture for massive star formation where the
qualitative physical processes are similar to those in low-mass star
formation, i.e., accretion processes mediated by accretion disks and
associated outflows/jets. The main differences are 1) the clustered
mode of high-mass star formation which produces observational problems
and the need for high-spatial resolution, and 2) that the most
relevant physical processes are quantitatively much stronger, i.e.,
higher accretion rates, higher outflow rates and masses, larger
luminosities, more UV radiation, etc. On a cautionary note, it should
be stressed that the observed molecular outflows rarely exceed regions
where the central objects are more massive than 30\,M$_{\odot}$.
Therefore, going to higher-mass regions, we may encounter more
severe changes than so far observed.

\begin{figure}[htb] \centering
\caption{The
left and middle panel present PdBI observations toward the young
massive star-forming region IRAS\,05358+3543 \cite{beuther2002d}. The
grey-scale shows the shocked H$_2$ emission and the blue/red contours
the CO(1--0) and SiO(2--1) emission, respectively. The arrows and large
ellipses guide the eye for the various outflow directions, and the
pentagons show additional H$^{13}$CO$^+$(1--0) peak positions. The
top-left inlay zooms into the marker box with high-velocity CO
emission in contours and 3\,mm continuum in grey-scale. The right
panel presents the corresponding 1.2\,mm continuum (grey-scale)
and blue/red CO(2--1) emission (contours) observed previously with the
IRAM\,30m single-dish telescope.}  \label{05358} 
\end{figure}

\section{Accretion disks}
\label{disks}

Angular momentum conservation and rotation of molecular cores always
predicted that during star formation gas and dust has to assemble in
disk-like structures around the forming central protostar. However,
evidence was lacking for a long time. First observational indications
for the existence of accretion disks was based on single-dish
multi-wavelengths observations of TTauri stars. For example, analyzing
the spectral energy distributions of a sample of low-mass protostars
from optical to mm wavelength, the mm dust continuum emission implied
such high column densities that in spherical symmetry one would not
have detected the sources in the optical or infrared
\cite{beckwith1990}.  However, since these sources were optical
visible TTauri stars, spherical symmetry was excluded, and
disk-symmetry appeared the most likely way to solve this problem. Then
in the mid-90th, Hubble Space Telescope observations revealed
accretion disks in Orion as absorption shadows against the strong
background radiation (e.g., \cite{mccaughrean1996}).

Since accretion disks are dense, flattened dust and gas condensations,
they have on average relatively low temperatures of the order a few
10\,K. Therefore, if one wants to learn more about their structure,
dynamics and kinematics, the mm and submm bands are the wavelength
regimes of choice. Furthermore, with disk sizes of the order 100\,AU
at typical Taurus distances of 150\,pc, a spatial resolution of $\leq
1''$ is necessary to resolve any substructure. Although disks in
massive star formation are likely larger of the order 1000\,AU, their
average distance is $\geq 2$\,kpc, and again one needs sub-arcsecond
spatial resolution to study such objects. Hence (sub)mm interferometry
is the tool of choice.

Observational studies have focused over recent years on several issues
in accretion disk studies, among them are 1) dust evolution, the
formation of larger grains and the way to planet formation, 2)
kinematic studies of disks and their stability, 3) chemical properties
and chemical complexity in accretion disks. While chemistry will be
discussed in \S\ref{chemistry}, important results for topics 1)
and 2) will be summarized here.

It is known for centuries that the planets of our solar system are
located approximately within the ecliptic plane originating from the
initial accretion disk, and that their velocity structure can be well
explained by the Keplerian laws. Because the formation of planets is
so tightly linked with accretion disks, dust evolution and the
formation of larger objects is essential for our understanding of the
earth history. One way to study dust properties is to observe
accretion disks at different wavelength and then investigate their
spectral energy distribution. In the Rayleigh-Jeans limit, which is a
good approximation at mm wavelength, the mm continuum flux $S$ scales
with $\nu^{2+\beta}$, where $\nu$ is the frequency and $\beta$ the
dust opacity index ($\tau\propto \nu^{\beta}$). For normal
interstellar grains $\beta$ is $\sim 2$ \cite{hildebrand1983}. With
increasing grain sizes this index $\beta$ decreases continuously.
Single-dish studies of a sample of accretion disk candidates found
lower values of $\beta$ and claimed already grain growth in these
sources \cite{beckwith1990}.  However, with single-dish studies alone,
the results are ambiguous because a decreasing spectral index can also
be mimicked by an increasing optical depth within the disk. Only
spatially resolved interferometric disk studies later could determine
the disk density structure, infer their optical depth and hence better
determine that grain growth actually takes place in accretion disks
\cite{beckwith2000,natta2004}. Incorporating more sophisticated dust
and radiative transfer models, one can infer density distributions,
temperature distributions and dust composition in even better detail
(e.g., \cite{calvet2002,wolf2003,natta2004,schegerer2006}).

An interesting new way to study dust and disk properties is
interferometry at mid-infrared wavelength.  The Very Large Telescope
Interferometer (VLTI) offers the MIDI instrument \cite{leinert2004}
which allows two-baseline interferometry at wavelengths between 8 and
12\,$\mu$m covering a strong silicate band. In one such study, it was
found that the inner disk regions has a higher degree of
crystallization than usually observed in the interstellar medium
\cite{vanboekel2004}.

As the evolutionary next step, the dust around debris disks can be
studied at (sub)mm wavelengths as well. For example, Wilner et
al. \cite{wilner2002} observed the 1.3\,mm dust continuum emission
toward the Vega debris disk, and they found two dust condensations
between 60 and 100\,AU from the exiting star. Such a dust distribution
can be explained by the dynamical influence of an unseen planet of a
few Jupiter masses in a highly eccentric orbit that traps dust in
principal mean motion resonances \cite{wilner2002}.

\begin{figure}[htb]
\centering
\includegraphics[width=11.5cm]{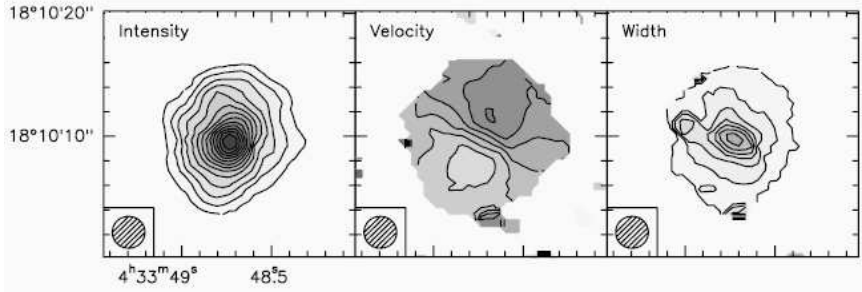}
\caption{CO(1--0) emission from the Keplerian protostellar disk DM
Tauri \cite{guilloteau1998}. The left panel shows the integrated
intensity, the middle panel the velocity field and the right panel the
line width.}
\label{gmtau}  
\end{figure}

To investigate disk kinematic properties, one needs spectral line
observations of the disks. The best accessible target for such kind of
studies are TTauri stars in the relatively evolved class II stage of
protostellar evolution \cite{andre2000}. At this stage, most of the
original protostellar envelope has already been dispersed, and the
accretion disks remains rather isolated and undisturbed for
observational studies (In younger regions, one suffers from confusion
between disk and envelope contributions to the observed spectral
lines.). A good example of CO emission from a protostellar disk is the
study of DM Tau with the PdBI (Fig.~\ref{gmtau},
\cite{guilloteau1998}). They resolve the velocity structure of the
disk finding a velocity distribution along the disk axis consistent
with Keplerian rotation (i.e., the centrifugal force equals the
gravitational force $F_{\rm{cen}} = \frac{mv^2}{r} = F_{\rm{g}} =
\frac{Gm_*m}{r^2} \rightarrow v = \sqrt{\frac{Gm_*}{r}}$). This also
implies that the observed linewidth increases getting closer to the
central star (see Fig.~\ref{gmtau}). In a more statistical sense,
larger samples of accretion disk sources were observed at high spatial
resolution, and typical Keplerian velocity structure was found in many
objects (e.g., \cite{simon2000}).

However, also different disk structures have been observed. For
example, the disk around the prototypical Herbig Ae star AB Aurigae
shows a central depression in the cold dust and gas emission and a
non-Keplerian velocity profile ($v=r^{-0.4\pm 0.01}$)
\cite{pietu2005,lin2006}. Possible explanations for such deviations
from more typical disk structures could be either the formation of a
low-mass companion or planet in the inner disk, or a much earlier
evolutionary phase where the Keplerian motion is not yet
established. Hence, non-Keplerian velocity structures are expected for
very young as well as more evolved accretion disks.

While the picture for low-mass accretion disks has tremendously
evolved over the last decade, progress in massive disk studies has
been significantly slower. This is again partly due to the on average
larger distances of massive star-forming regions, and due to the
clustered mode of massive star formation that complicates the spatial
isolation of massive accretion disks. However, an additional problem
arises from the fact that massive star formation proceeds much faster
than low-mass star formation (of the order $10^5$\,yr compared to a
few times $10^6$\,years), and that evolved massive stars will
relatively quickly dissipate the original accretion disks. Therefore,
it is unlikely to find a massive accretion disk in a similarly evolved
and exposed state like the low-mass TTauri stars, but we have to
search for these objects in the much younger and deeply embedded
evolutionary stages. This implies, that molecular line as well as dust
continuum emission is usually not solely attributable to an accretion
disk, but that we always have additional contributions from the
larger-scale core and envelope emission. In many cases, this envelope
emission can dominate any line and continuum study. One technical
advantage of interferometers is that they filter out the emission on
large spatial scales because they are not sensitive to structures
$\theta$ corresponding to baseline lengths $D$ below the shortest
separation of two baselines ($\theta\sim\frac{\lambda}{D}$, for more
details see, e.g., \cite{taylor1999,thompson2001}). Therefore, we can
filter out large parts of the envelope emission, nevertheless, not all
emission is usually filtered out, and one still suffers from confusion
between genuine disk and surrounding core emission. To overcome this
difficulties, one is interested to observe molecular lines that are
usually weak or not found in the envelope, but that should be strong
in the inner disk regions. Typical hot core molecules like CH$_3$CN,
HCOOCH$_3$, C$^{34}$S, HN$^{13}$C or torsionally excited CH$_3$OH
transitions are good candidates for such lines. However, observations over
the last few years have shown that in many sources exclusively one or
the other tracer only allows rotational studies, whereas other
candidate spectral lines are either too weak, optically thick or maybe
have not been formed in the chemical network yet. For more details on
these problems see a summary in a recent review
\cite{beuther2006d}. This difficulties complicate statistical studies
of massive accretion disks, and our current knowledge is based on only
a few selected examples.

One of the best known massive disk candidates is within the high-mass
star-forming region IRAS\,20126+4104
\cite{cesaroni1997,cesaroni1999,zhang1998a,cesaroni2005}. Observations
in various spectral lines clearly identify a velocity gradient
perpendicular to the molecular outflow/jet, and the velocity structure
is consistent with Keplerian rotation \cite{cesaroni2005}. Based on
the Keplerian motion, the mass of the central object is estimated to
"only" 7\,M$_{\odot}$, hence a large fraction of the observed
luminosity of the order $10^4$\,L$_{\odot}$ has to be due to accretion
luminosity \cite{krumholz2006b}. Based on the existing data, it is
suggested that this source is still actively accreting and may well
form a much more massive star at the of its evolution
\cite{cesaroni2005}. For most other massive star-forming regions of
higher luminosity, no clear signatures of Keplerian rotation have been
found so far. The current interpretation of larger-scale rotation
signatures is that such larger-scale entities with sizes of a few
1000\,AU rather resemble rotating, probably infalling toroid that may
harbor more genuine accretion disk at their so far unresolved centers
\cite{cesaroni2006}. The likely most extreme case of rotating and
infalling signatures has been observed toward the well-known very
luminous ($\sim 10^6$\,L$_{\odot}$) ultracompact H{\sc ii} region
G10.6. Rotational infalling signatures were observed on larger scales
in the molecular gas (NH$_3$ lines,\cite{sollins2005}), and on much
smaller scales similar signatures were identified in the ionized gas
(H$_{\alpha}$ recombination line, \cite{keto2002a}). These
observations likely trace a larger-scale in-spiraling structure that
continues from the molecular to the ionized gas. In classical H{\sc
  ii} region theory that would not have been possible because the
formed H{\sc ii} region should expand by the pressure and not allow
any further infall and accretion. The observations of G10.6 allow a
different scenario, where the gravity still dominates the structure of
the UCH{\sc ii} region (not yet the pressure), and hence accretion
through the evolving H{\sc ii} regions may be possible for some time
\cite{keto2002b,keto2003}.

\begin{figure}[htb]
\centering
\includegraphics[width=5.7cm]{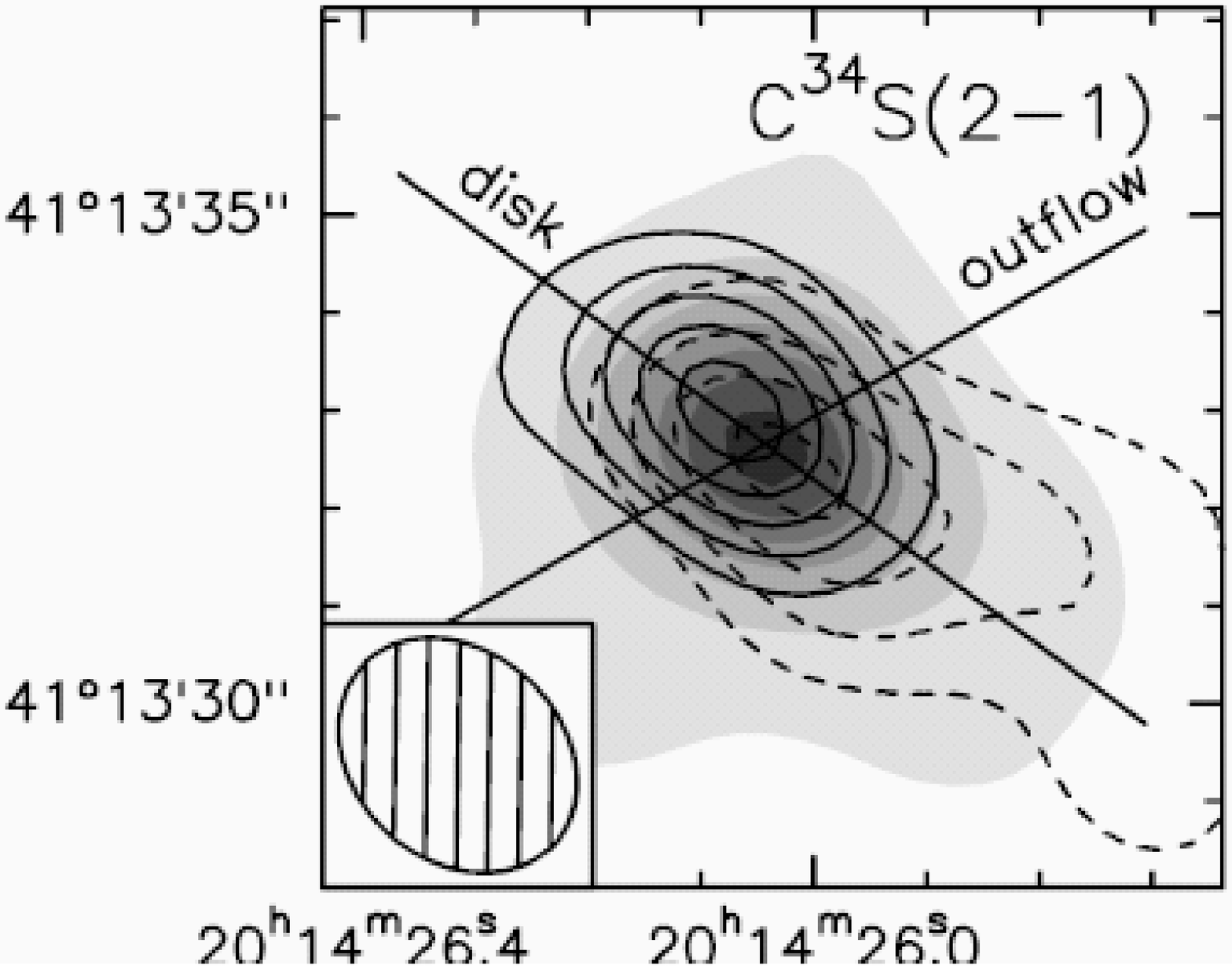}
\includegraphics[width=5.5cm]{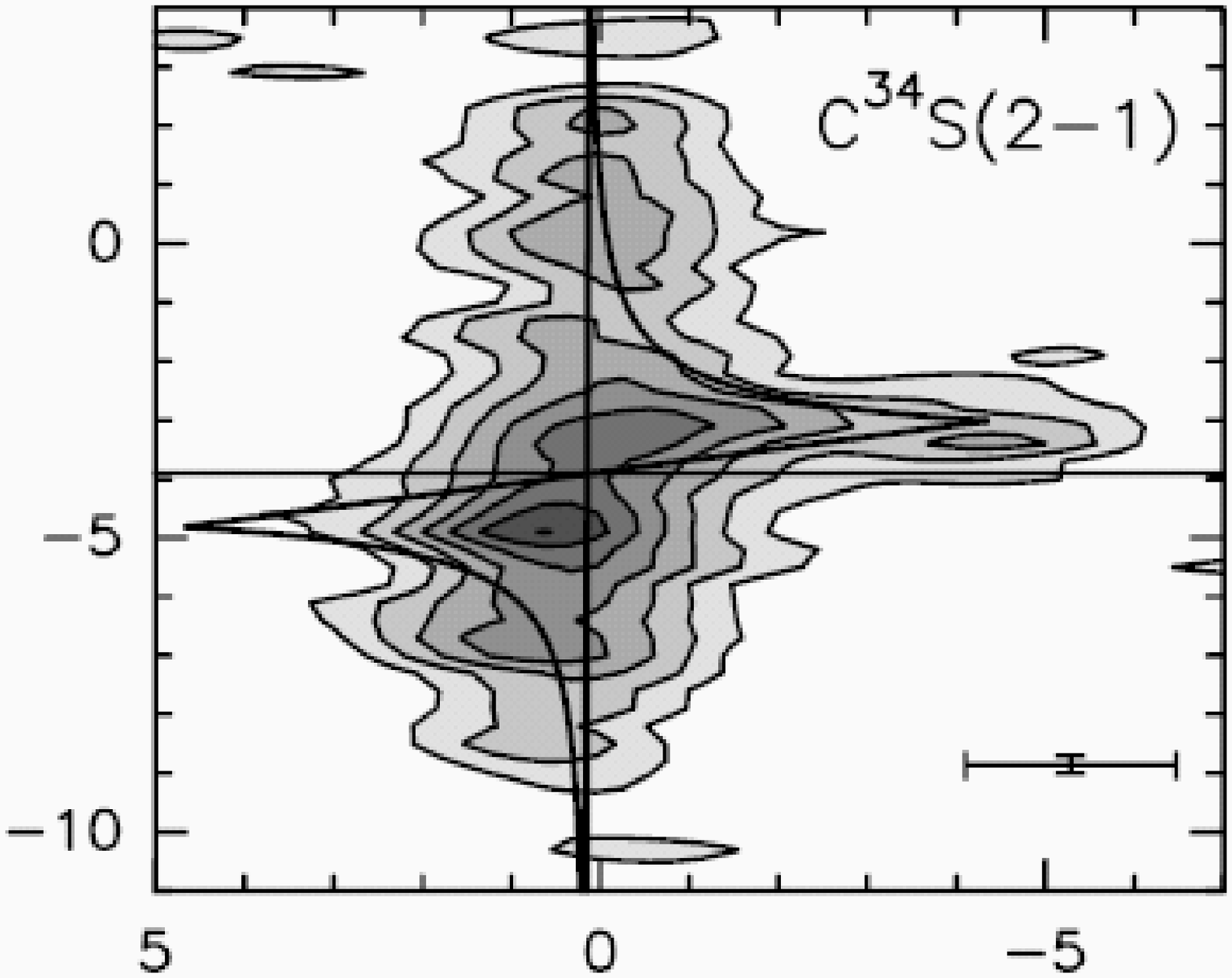}
\caption{Keplerian disk emission toward the HMPO IRAS\,20126+4104
\cite{cesaroni2005}. The left panel shows in grey-scale the 3\,mm
continuum emission and in full and dashed contours the blue- and
red-shifted C$^{34}$S(2--1) emission. The outflow and disk axis are
indicated. The right panel presents a position velocity diagram (X-axis
offset, Y-axis velocity). The full line shows the expected emission of
a Keplerian disk.}
\label{20126}  
\end{figure}

\section{Fragmentation}
\label{fragmentation}

Since its first determination in 1955 \cite{salpeter1955}, the rather
universal validity of the Initial Mass Function (IMF) has been
confirmed in many ways. For a recent compilation see
\cite{corbelli2005}. One of the essential question in general star
formation research is how and at what evolutionary stage the shape of
the IMF gets established. Turbulent fragmentation theories predict that
the turbulence right at the onset of fragmentation processes already
produces core mass functions with the same power law like the IMF, and
that by applying star formation efficiencies these core mass functions
directly convert into the later observed IMF
\cite{padoan2002,maclow2004}. Contrary to that, other groups argue that
the the early star-forming gas clumps fragment down to many cores of
more or less the Jeans mass ($\sim 0.5$\,M$_{\odot}$), and that the
IMF will be formed from that stage via competitive accretion from
previously unbound gas \cite{bonnell2004,bonnell2006}.

\begin{figure}[htb] \centering
\includegraphics[angle=-90,width=11.7cm]{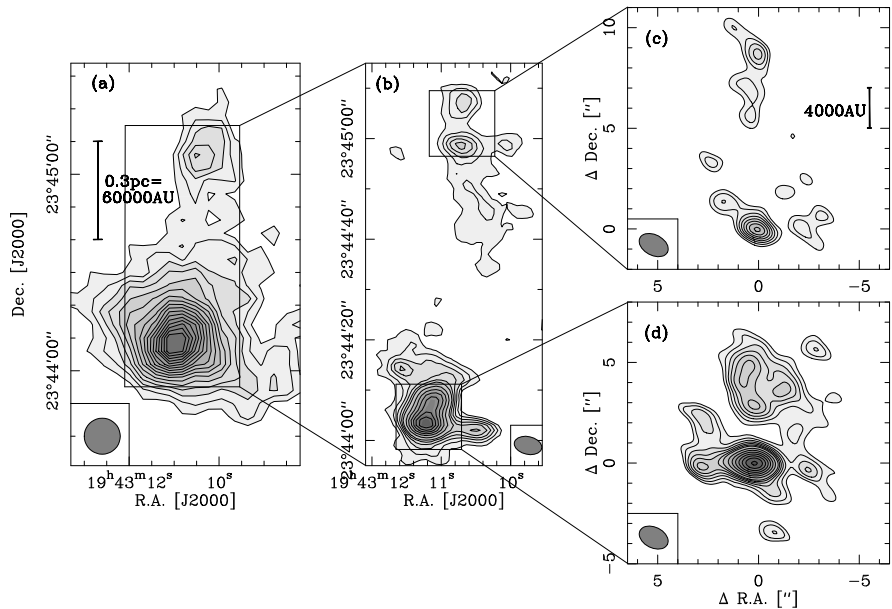}
\caption{Single-dish and interferometer mm dust continuum observations
toward the young high-mass star-forming region IRAS\,19410+2336
\cite{beuther2004c}. The left panel shows the 1.2\,mm map obtained
with the IRAM 30\,m telescope. The middle and right panel present 3
and 1.3\,mm continuum images from the PdBI with increasing spatial
resolution. The synthesized beams are shown at the bottom-left of each
panel. The resolution difference between the left and right panels are
an order of magnitude.} \label{19410} \end{figure}

In the low-mass regime, dust continuum studies with bolometer arrays
installed on single-dish telescopes revealed the core mass
distributions in nearby regions like $\rho$ Ophiuchus and Orion. The
power-law distributions $dN/dM \propto M^{-\alpha}$ of the pre-stellar
dust condensations resembles the Salpeter IMF with typical values of
$\alpha$ between 2 and 2.5. While this finding is support for the
turbulent fragmentation scenario, it only covers a small range of the
IMF, and it is necessary to expand such studies to higher-mass
regions. Several groups investigated with the same single-dish
instruments samples of massive star-forming regions, and the derived
cumulative mass distributions of their samples resemble the Salpeter
IMF as well
\cite{shirley2003,williams2004,reid2005,beltran2006}. However, since
these massive star-forming regions are at about an order of magnitude
larger distances, these single-dish studies do not resolve individual
protostars but they average over the whole forming protocluster. Hence,
these cumulative mass distributions rather resemble protocluster mass
functions, and it is far from clear whether they can set any
constraints on the formation of the IMF. Therefore, again high spatial
resolution is required to resolve individual protostars within the
forming protoclusters. So far, only one study resolved and imaged
enough sub-sources within a massive star-forming regions that a
derivation of a core mass function appeared meaningful. The two
massive gas condensation in IRAS\,19410+3543 were resolved in 1.3\,mm
continuum observations with the PdBI into 24 sub-sources
(Fig.~\ref{19410}, \cite{beuther2004c}), and the resulting core mass
function again has a power-law distribution with $\alpha\sim
2.5$. Although the statistics are still poor, and a few caveats have
to be taken into account in this analysis (Are all dust continuum
peaks of protostellar nature? Is the assumption of the same
temperature for all sub-sources justified?), it is exciting that all
studies so far find core mass distributions resembling the IMF,
indicating that turbulent fragmentation processes really are
important. However, especially in the high-mass regime, we cannot draw
any definitive conclusions yet. Larger source samples observed at high
angular resolution are required to base the results on solid
grounds. Furthermore, it will be necessary to investigate the
temperature structure of the sub-sources in detail, and we need
additional spectral line data to determine complementary virial masses
and thus establish that the observed sub-sources are really bound and
not transient structures. 

\begin{figure}[htb] \centering
\includegraphics[angle=-90,width=6.1cm]{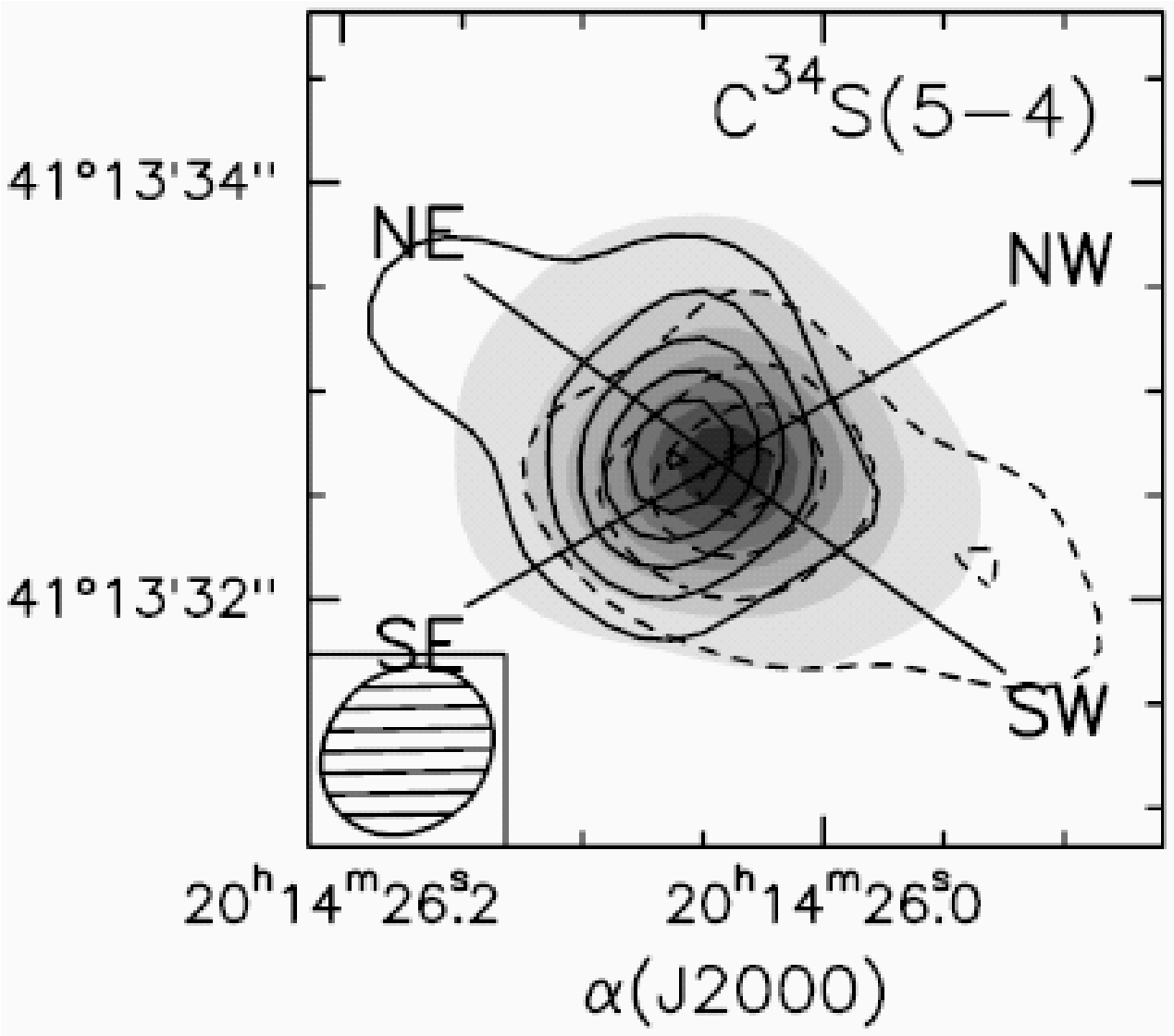}\
\includegraphics[angle=-90,width=5.3cm]{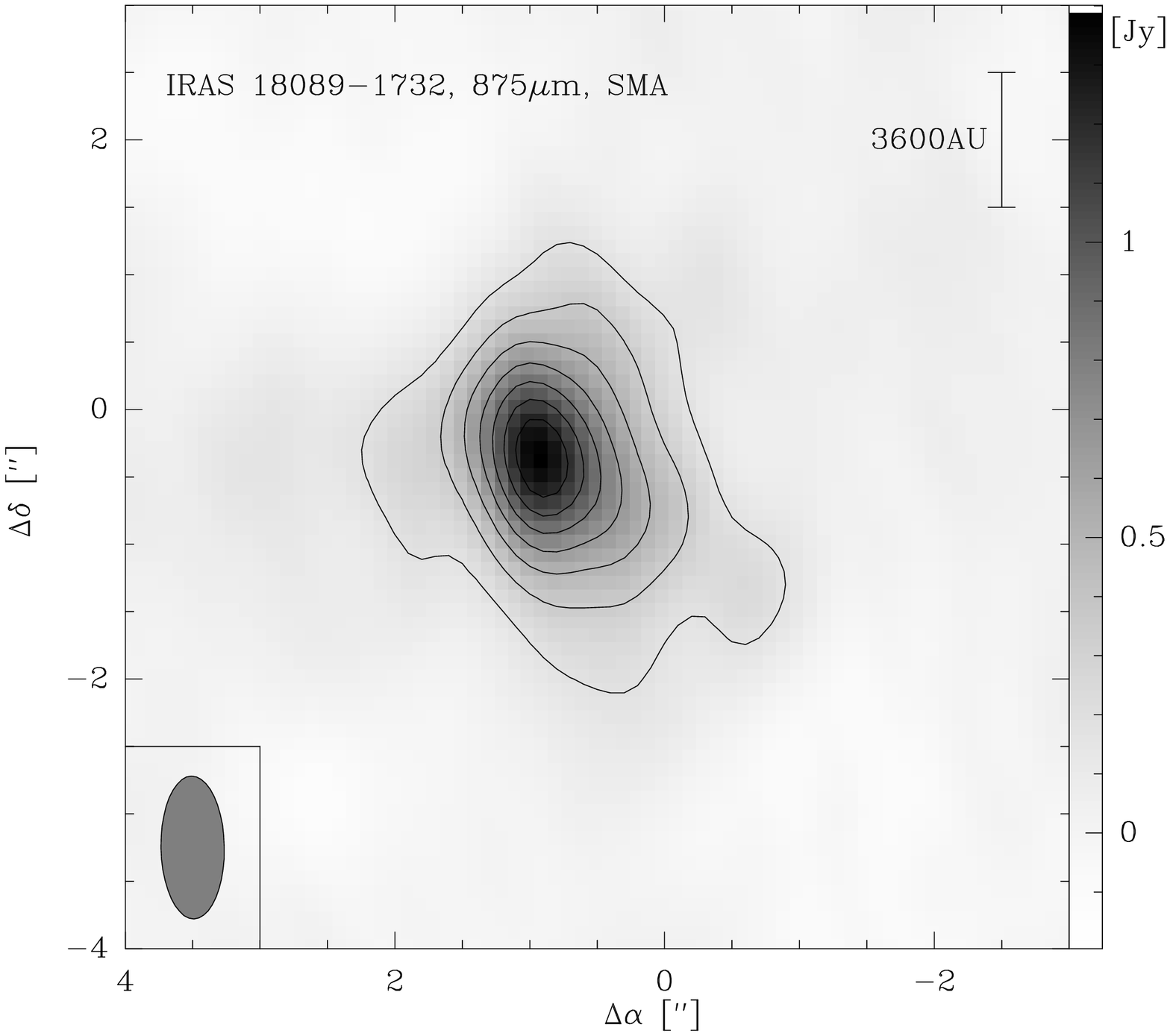}\
\includegraphics[angle=-90,width=6.1cm]{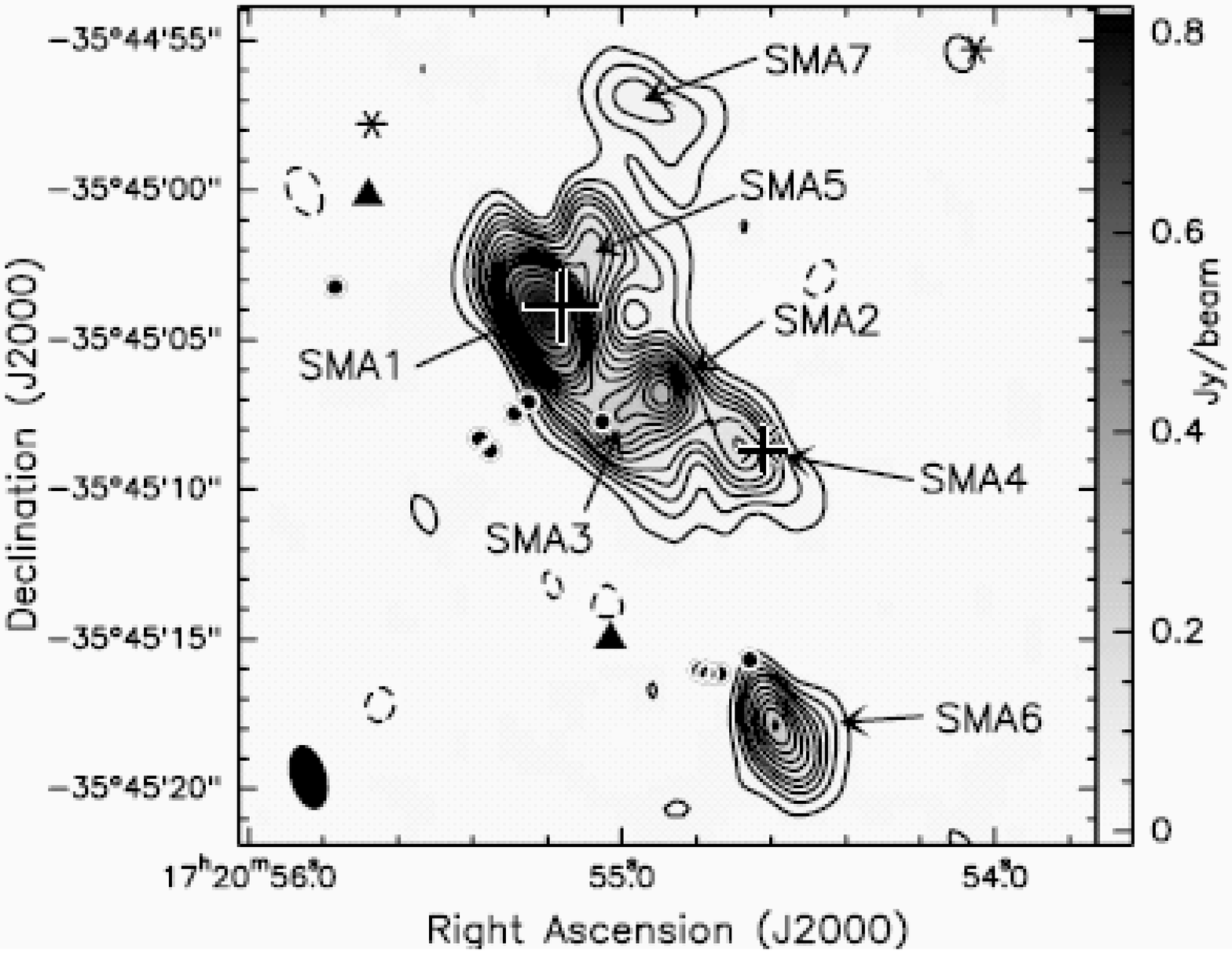}
\includegraphics[angle=-90,width=5.3cm]{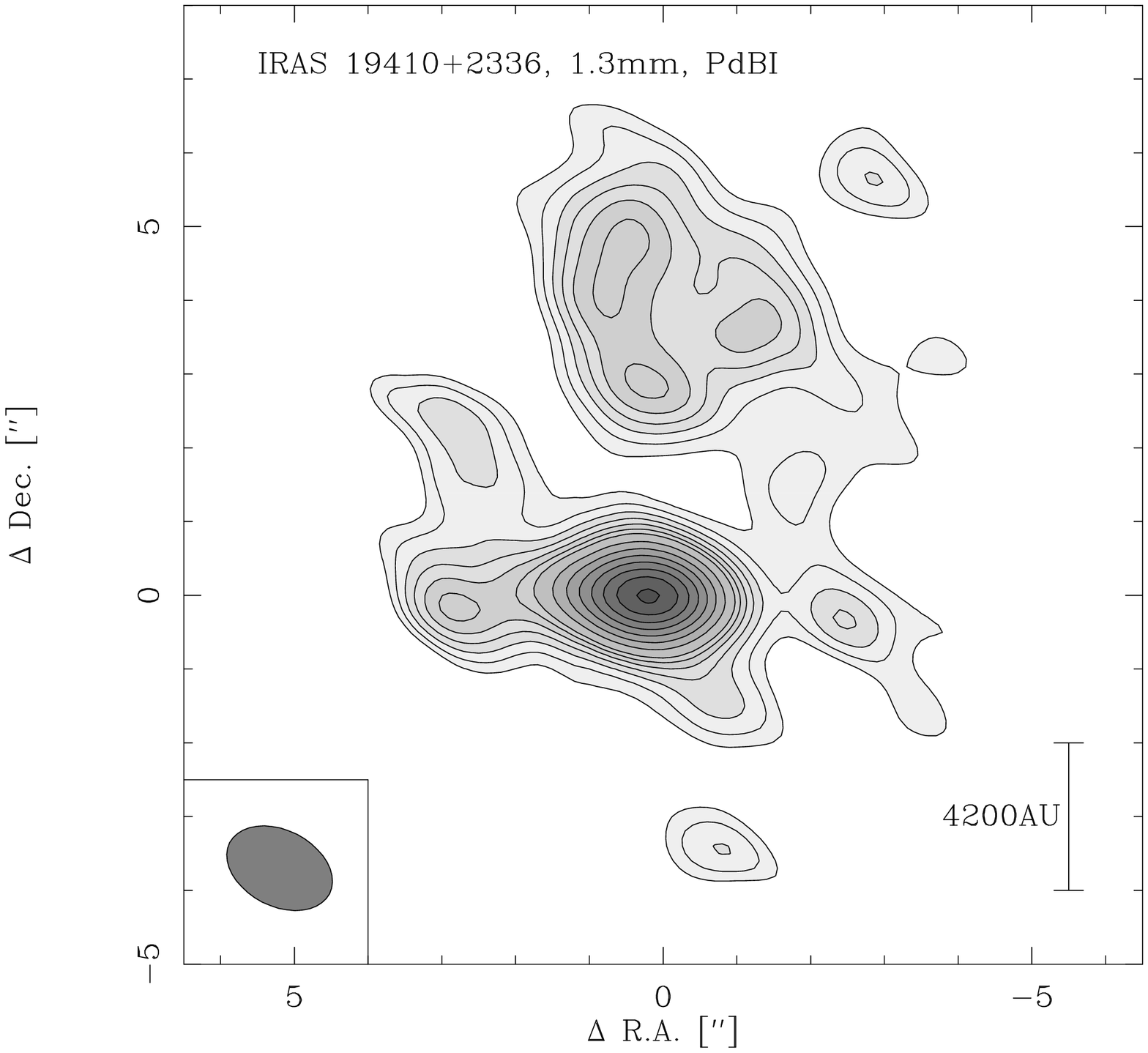}
\caption{Interferometric high-spatial-resolution (sub)mm dust
continuum images of sources at comparable evolutionary stage and
luminosity but with very different fragmentation characteristics. The
top-left panel shows IRAS\,20126+4104 (grey: 1.3\,mm continuum,
contours C$^{34}$S(5-2), 1.7\,kpc, \cite{cesaroni2005}), the top-right
is IRAS\,18089-1732 (875\,$\mu$m, 3.6\,kpc, \cite{beuther2005c}), the
bottom-left is NGC6334I(N) (1.3\,mm, 1.7\,kpc, \cite{hunter2006}, the
bottom-left is IRAS\,19410+2336 (1.3\,mm, 2.1\,kpc,
\cite{beuther2004c}). The synthesized beams are shown at the
bottom-left of each image.} \label{sample} \end{figure}

Another observational curiosity of several high-mass star-forming
regions, that appear very similar from previous single-dish
observations (They are at similar distances, should be at
approximately the same evolutionary stage, and have comparable
luminosities and other massive star-forming tracers like masers and
outflows.), is that they reveal very different sub-structures when
observed at high angular resolution. Figure \ref{sample} shows a few
interferometric example studies of regions where one would have
expected comparable fragmentation results. However, some of the
regions show only a single massive central source, even at the highest
angular resolution (Fig.~\ref{sample} top row), whereas other sources
exhibit many sub-sources as expected from a star-forming cluster
(Fig.~\ref{sample} bottom row). While the latter is less of a surprise
since we know that massive stars almost always form in clusters,
finding only a single peak in other regions is more difficult to
explain. While all these studies are sensitivity limited and usually
are not capable to trace any condensation below approximately
1\,M$_{\odot}$, one would nevertheless expect to find at least some
intermediate-mass objects in the vicinity of the central object. One
important additional fact is that toward most regions deeply embedded
near-infrared clusters have been found, and hence they are no isolated
objects. The existence of an embedded near-infrared protocluster in
the vicinity of a massive forming star may be interpreted in the
direction that the low-mass stars form first and the high-mass objects
later (see also \cite{kumar2006}). However, currently we do not
understand why various apparently similar massive star-forming regions
show such a diverse fragmentation behavior. Is it only an
observational bias and the selected sources may be in fact not as
similar as we believe, or may there be different paths massive
star-forming regions can fragment in the first place?

\section{Chemistry}
\label{chemistry}

Astrochemistry is a steadily rising topic over recent
years. Large-scale chemical mapping surveys of molecular clouds (e.g.,
\cite{bergin1997,ungerechts1997} and line surveys toward high-mass hot
molecular cores \cite{schilke1997b,hatchell1998b}, both conducted with
single-dish instruments, have already been conducted for more than two
decades. However, studying the chemical small-scale diversity of hot
molecular cores has been a difficult task because the spectral
bandpasses available for most existing interferometers were mostly too
small to cover enough molecular lines from various species in a
reasonable amount of time (a noteworthy exception is
\cite{blake1996}). This has changed significantly since the advent of
the SMA with its two times 2\,GHz bandwidth in the upper and lower
sideband \cite{ho2004}, and similar capabilities to be available very
soon at the PdBI and CARMA. As a prime showcase of the spatial
chemical complexity of hot molecular cores, here a recent
investigation of Orion-KL will be presented.

The Orion-KL region at a distance of 450\,pc is the closest and 
prototypical hot molecular core. With the advent of the SMA, it
obviously became one of the early targets of this instrument, and the
region was mapped at arcsec spatial resolution in the 865\,$\mu$m
submm wavelength band \cite{beuther2005a}. The 4\,GHz bandpass covered
more than 150 spectral lines from at least 13 species, 6 isotopologues
and 5 vibrational excited states within a single observation run. From
a chemical point of view, the observations included nitrogen- as well
as oxygen-bearing species, shock-tracer like SiO and SO, more than 40
CH$_3$OH lines, and many complex molecules like CH$_3$CH$_2$CN or
HCOOCH$_3$. While this chemical diversity is interesting in itself, it
also allows to study various physical properties in detail, e.g., the
SiO data can be used for a kinematic analysis of the molecular
outflow(s) in the region, and the large number of CH$_3$OH lines
(ground state as well as torsionally excited $v_t=1,2$ lines) are an
ideal tool to investigate the temperature structure of the whole
region. Furthermore, the data also allowed to derived the first
sub-arcsecond resolution dust continuum map of this region
\cite{beuther2004g}, clearly differentiating the central power house
source {\it I} from the hot molecular core, detecting source {\it n} for
the first time in cold dust emission, and identifying a new
protostellar source SMA1, which may be the driver of one of the
outflows in this complex star-forming region.

\begin{figure}[ht] \centering
\caption{SMA
submm spectral line observations toward Orion-KL
\cite{beuther2005a}. The figure shows representative images from
various molecular species labeled in each panel. Full contours show
positive emission, dashed contours negative features due to missing
short spacings. The stars mark the locations of source {\it I}, the
hot core peak position, the newly identified source SMA1 and source
{\it n} (see bottom-right panel).}  \label{orion} \end{figure}

Figure \ref{orion} shows a small number of line images from the whole
dataset to outline the chemical complexity of the Orion-KL hot
molecular core. While SiO traces the collimated outflow structure
centered on source {\it I}, all other species show a more complex
morphology. Nitrogen-bearing species like CH$_3$CN or CH$_3$CH$_2$CN
exhibit the kind of horse-show morphology which was know for the hot
core already from early NH$_3$ observations (e.g., \cite{wilson2000}).
In contrast, the oxygen-bearing molecules, foremost CH$_3$OH, are
relatively speaking weaker there, but they show a strong additional
peak toward the south-west, the so-called compact ridge. This southern
region is believed to be the interface of one of the molecular
outflows (as traced in SiO) with the ambient cloud, and hence the
CH$_3$OH is strongly excited in this shocked interface region. Other
molecules, e.g.~C$^{34}$S, show a mixture of these morphologies and
sometimes even an additional peak positions toward the north-west
associated with the infrared-source IRC6. The main lesson one may take
home from this spatial molecular differences is that massive
star-forming regions are far more complex than one generally assumes,
and that one always has to take into account various physical and
chemical processes like heating, outflows, shocks or chemical
evolution if one wants to model such regions accurately. While
single-dish surveys give a good average molecular inventory of such
regions, they do not allow to set more detailed constraints.
Interferometric imaging surveys are needed to investigate their
structures in depth.

\section{Outlook and Future}

Interferometry at mm wavelength has been conducted for more than one
and a half decades by now, but only recently with the advent of the
SMA, the submm spectral window started to be accessible for
high-spatial-resolution observation. Both wavelength regimes can be
considered as close siblings where a range of scientific questions can
be targeted by both, but where also some other questions are unique for
each band. For example, hot molecular cores are by definition strong
in the submm regime whereas colder and younger regions are more easily
studied in the mm band. Therefore, none of the bands is obsolete, but
they complement each other well. The main facilities in the
northern hemisphere are and will be the SMA, the PdBI and CARMA, and
they will be the sites of many exciting science over the coming
decade. 

Nevertheless, the upcoming Atacama Large Millimeter Array (ALMA),
which will be built within the next few years in Chile by a worldwide
collaboration of North-America, Europe and Japan, will supersede the
already existing instruments in many ways by orders of magnitude. At
an altitude of $\sim$5000\,m, ALMA will consist of fifty 12\,m dishes
(25 provided by Europe and 25 by the USA), combined with a few 7\,m
antennas provided by the Japanese partners to supplement the short
spacing information (the Atacama Compact Array, ACA). The baselines
range will be from 15\,m and 15\,km, and the receivers are expected
to cover all earth-accessible spectral windows between 30 and
950\,GHz. The spectral first-light bands are expected to be around
100, 230, 345 and 690\,GHz, and the bandpass will be broad with 8\,GHz
and dual polarization capability. Combining these specs, the
anticipated 350\,GHz continuum sensitivity is 1.4\,mJy in 1 second
integration time. For more detailed descriptions of the specs and the
scientific capabilities of ALMA, one may visit
http://www.eso.org/projects/alma/.

The requirements on the ALMA capabilities have been driven by mainly
three scientific topics:\\ - Detect CO or CII in a normal galaxy at
z=3 in less than 24 hours ($\rightarrow$ sensitivity).\\ - Image
protostars and proto-planetary disks around sun-like stars at 150pc
(Taurus) to study kinematics, chemistry, magnetic fields and tidal
gaps created by forming planets ($\rightarrow$ spatial resolution down
to 10mas).\\ - Good imaging quality down to $0.1''$ resolution
($\rightarrow$ spatial resolution and uv-coverage/number of
antennas).

In the framework of the topics discussed in this chapter, a few likely
highlights should be mentioned. With the highest spatial resolution,
we will be able to probe the central launching regions of the outflow
and jets and hence observationally much better constrain the physical
processes governing these energetic processes of energy and momentum
transfer. As outlined above, disk/planet research has been one of the
main drivers for ALMA, and we will witness the planet-formation
processes within the young accretion disks (e.g., \cite{wolf2005}).
For massive accretion disk studies, the advent of ALMA will be
important in many ways: the high spatial resolution will help to
better disentangle the clustered sub-structure of the regions, the
even broader bandpasses will allow to sample many spectral lines
simultaneously and thus give a broad range of analysis tools within
single observations, and the high sensitivity will give access to the
weaker, optically thin and less abundant lines, which are likely the
best candidate lines to differentiate the genuine disk emission from
the surrounding envelopes. Regarding future fragmentation studies,
ALMA will allow to study larger source-samples in a consistent manner,
which is needed to draw statistically significant constraints.
Furthermore, the instantaneous extremely good uv-coverage provides an
order of magnitude better image fidelity, which is very important to
derive credible core-mass distributions in young massive star-forming
regions. Last but not least, spectral imaging surveys will gain
significant more momentum: Again based on the broad bandpasses and
high sensitivity, such surveys can be done is a far more consistent
manner for large source samples covering various evolutionary stages
and extending from the high-mass to the low-mass regime. This will
help to differentiate much better the chemical evolution depending on
mass as well as on time. Furthermore, combining the good uv-coverage
of the main ALMA array with the short spacing information from the
ACA, we will not have the problem of negative features caused by the
missing short spacings as shown in Fig.~\ref{orion}. These missing
flux problem so far severely hampered correct column density
determinations. Overcoming these difficulties is crucial for accurate
abundance determinations which are one of the important parameters in
any evolutionary chemical network. In addition, it is expected that
spectral imaging surveys will greatly help to identify new molecules
and thus potentially pave the way to find the first bio-molecules in
space.

Although this chapter could only roughly outline the potential of
(sub)mm interferometry, it has to be stressed that this field is
extremely vibrant and, many exciting questions are waiting to be
investigated. The progress over the last two decades has been
enormous, and we currently have a number of excellent facilities to
study many important astrophysical/astrochemical questions. In
addition to this, the future is bright with ALMA on the horizon. This
new facility will add another quantum leap to (sub)mm interferometry
and probably nearly all fields of astrophysical research.\\

H.B.~acknowledges financial support by the Emmy-Noether-Program of the
Deutsche Forschungsgemeinschaft (DFG, grant BE2578).



%


\end{document}